**High frequency study of the orbital ordering resonance in the strongly correlated heavy fermion metal CeB$_6$.**

S.V. Demishev[a], A.V. Semeno[a], H. Ohta[b], S. Okubo[b], Yu.B. Paderno[c], N.Yu. Shitsevalova[c], N.E. Sluchanko[a]

[a]*Low Temperatures and Cryogenic Engineering Department, A.M.Prokhorov General Physics Institute of Russian Academy of Sciences, Vavilov street, 38, 119991 Moscow, Russia*

[b]*Molecular Photoscience Research Center, Kobe University, 1-1 Rokkodai, Nada, Kobe 657-8501, Japan*

[c]*Department of Refractory Materials, Institute for Problems of Materials Science of Ukranian National Academy of Sciences, Krzhyzhanovsky street, 3, 03680 Kiev, Ukraine*

Short title: Orbital ordering resonance in CeB$_6$.

Communicating author: Professor Sergey Vasil'evich Demishev

        Low Temperatures and Cryogenic Engineering Department,

        A.M.Prokhorov General Physics Institute of Russian Academy of Sciences,

        Vavilov street, 38

        119991 Moscow, Russia

        Tel./Fax: +7-495-1358129

        E-mail: demis@lt.gpi.ru

**Abstract.**

We report results of the study of the recently discovered magnetic resonance in the orbitally ordered phase of $CeB_6$ (the orbital ordering resonance) in a wide frequency range $\omega/2\pi$=44-360 GHz. It is found that the *g*-factor for this resonance increases with frequency from $g(\omega/2\pi=44\text{ GHz})\sim1.55$ to $g(\omega/2\pi>250\text{ GHz})\sim1.7$. In addition to the orbital ordering resonance for the frequencies exceeding 200 GHz a new magnetic resonance with the *g*-factor 1.2-1.3 is detected.

**1. Introduction.**

A strongly correlated heavy fermion metal $CeB_6$ is believed to be a classical example of a dense Kondo system, where $Ce^{3+}$ magnetic ions are arranged in the simple cubic lattice [1-7]. The magnetic phase diagram of this material is formed by a complicated interplay between spin and orbital degrees of freedom leading in zero magnetic field to the orbital ordering at $T_Q$=3.2 K, which precedes the formation of the long-range antiferromagnetic order at $T_D$=2.3 K. The application of the magnetic field $B$ induces an enhancement of $T_Q$ and suppression of $T_D$ [1-7]. In experiments this sequence of phase transitions in $CeB_6$ has been established by means of neutron diffraction [1] and resonant X-ray scattering [2] studies as well as by specific heat [3], NMR [4], magnetisation [5-6] and transport measurements [7].

The existing theoretical models describing physical properties and magnetic phase diagram of $CeB_6$ [8-14] exploit the idea of the interaction between quadrupole electric moments of the 4$f$ shells of Ce ions, which originate from the crystal field splitting of the $^2F_{5/2}$ level leading to the lowest in energy $\Gamma_8$ term [8-14]. The orbital ordering at phase boundary $T_Q(B)$ occurs without change of the lattice constant [2] due to the weak contribution of 4$f$ electrons to the chemical bounding, and, moreover, without change of the magnetic structure, which remains the same as in the paramagnetic phase at $T>T_Q(B)$ [8-15]. In $CeB_6$ the neutron scattering evidence [1] of an antiferromagnetic component with a wave vector $k_0$=[½, ½, ½] in the orbitally ordered phase suggests a theoretical consideration of the two types of non-equivalent Ce ions having at $T<T_Q$ quadrupole moments $+Q$ and $-Q$ and arranged in an alternating three-dimensional structure. Thus the orbitally ordered phase of $CeB_6$ at $T<T_Q$ is referred as an antiferro-quadrupole (AFQ) phase [1-15]. This physical situation is apparently different from the case of manganites, where the orbital ordering changes both crystal structure and magnetic order [16-21].

The aforementioned approach to genesis of the $CeB_6$ characteristics has been affected by a recent observation of the new magnetic resonance specific to the AFQ phase [22-24]. It is found that below the orbital ordering phase boundary a new, missing at $T>T_Q$, EPR-like mode with the *g*-factor $g=1.62$ develops [22]. Special efforts were made to show unambiguously that this mode reflects a bulk response of a strongly correlated matrix of $CeB_6$ rather than a surface or an impurity effect [22].

Initially for this magnetic oscillation phenomenon a term "antiferro-quadrupole resonance" have been suggested. However, the subsequent analysis of the static magnetic properties and magnetic resonance parameters carried out in [23] suggested that for $T<T_Q$ the magnetisation consists at least of two components, one of which is responsible for the observed magnetic oscillations and is missing in the paramagnetic phase [23]. The "oscillating" magnetic subsystem can be described as a set of the non-interacting magnetic dipoles with the magnitude $\sim 0.8\mu_B$ having concentration close to the concentration of $Ce^{3+}$ ions [23]. Apparently, in the quadrupolar ordering concept, where magnetism of Ce magnetic ions is solely accounted [8-15], it is difficult to explain the splitting of magnetisation into several components having different physical nature and interpret an appearance of the specific magnetic resonance in the orbitally ordered phase. Moreover, in a dense Kondo system the free magnetic moments necessary for a magnetic resonance should freeze out rather than emerge. Therefore the term for the discovered in [22-24] magnetic resonance based on the particular model of the orbital ordering may be misleading and below we will refer to it as to the orbital ordering resonance.

In any case, the observation of the new resonant mode in $CeB_6$ [22-24] puts on the agenda an adequate theory explaining magnetic properties of $CeB_6$ including magnetic resonance and other physical effects supposed to be caused by the orbital ordering. In the absence of such a theory it is worth to examine magnetic resonance in this material in more detail. Up to now only data for the cavity measurements at frequency $\omega/2\pi=60$ GHz were reported [22-24]. In the present paper we are aimed on the studying of the orbital ordering resonance in $CeB_6$ in a wide frequency range up to $\omega/2\pi \sim 360$ GHz.

**2. Experimental details.**

The frequency measurements have been carried out by using two different techniques. When studying a magnetic resonance in a low frequency range we have measured the transmission of the copper cylindrical cavities operating at $TE_{01n}$ modes and tuned to the frequencies 44, 60, 78 and 99 GHz. As a source of the microwave radiation a backward wave oscillators have been used. One of the endplates of the cylinder has been made of the high quality $CeB_6$ single crystal, and the holes connecting cavity to waveguides were located at the other endplate. The cavity quality factor was about $\sim 6 \cdot 10^3 - 2 \cdot 10^4$. As a reference a small DPPH sample has been placed in the cavity. The magnetic field was generated by the 7 T superconducting magnet.

In the high frequency range 100-360 GHz the quasi-optical schema for reflection measurements has been applied (fig. 1). To split the indenting and reflected beams a $45^o$ cooper mirror with a hole located inside oversize waveguide has been used. The $CeB_6$ single crystal was placed in the centre of the 30 T pulsed field magnet. For a reference purposes a thin layer of DPPH powder has been put on the top of the $CeB_6$ reflecting surface (fig. 1). Fast InSb detector has registered the reflected radiation.

In our experiments the magnetic field was aligned along the [110] crystallographic direction. The resonant magnetoabsorption spectra were measured at two temperatures 1.8 K and 4.2 K and all magnetic resonance data reported below corresponded to the orbitally ordered AFQ phase.

The quality of the $CeB_6$ single crystal has been controlled by means of X-ray and chemical analysis as well as by transport and magnetic measurements. The detail description of the sample surface preparation for the cavity and reflection measurements is provided elsewhere [22].

**3. Experimental spectra.**

The experimental spectra in cavity experiments at $T =1.8$ K are shown in fig. 2. It is visible that in the frequency range 44-99 GHz the single line A, which shifts almost linearly with magnetic field, is observed. The estimated $g$-factor value for the resonance A is about 1.6 in agreement with the previous findings [22-24]; in more detail the $g$-factors will be discussed below. Another characteristic aspect of the obtained data is the frequency independent kink in the vicinity of $B \sim 1$ T (fig. 2). The position of this feature exactly coincide with the phase boundary between antiferromagnetic (AFM) and antiferro-quadrupole phases at $T =1.8$ K, which is known from the literature [1-7] (dashed line in fig. 2). Therefore the studied magnetic resonance A in fact occurs in the orbitally ordered AFQ phase.

In the high frequency range, where reflection has been measured, the experimental spectra acquire a more complicated form. The typical results are presented in fig. 3, where for convenience of comparison the scale of the magnetic field is recalculated to the $g$-factor scale $g\sim\omega/B$. In addition to the expected resonance A with $g$-factor $\sim 1.6$ a new line B corresponding to the $g$-factors 1.2-1.3 has been observed for the frequencies exceeding $\omega/2\pi\sim 200$ GHz (fig. 3). Apparently no resonance B has been found at any frequency in the cavity experiment up to 7 T (fig. 2). However, similarly to the previous cavity experiments [22-24] in the pulsed field reflection measurements at $T=4.2$ K we have observed a kink at 1.6 T corresponding to the transition from the paramagnetic to the antiferro-quadrupole phase [1-7]. Thus, in our opinion, both resonances A and B reflect the high frequency properties of the orbitally ordered phase in $CeB_6$.

## 4. Discussion.

The data shown in fig. 3,4 were used to calculate the *g*-factor values, which are plotted in fig. 4 as a function of frequency. For the mode A the results of cavity and quasi-optical experiment agrees reasonably; however the data subtracted from the reflectance spectra at $T$ =1.8 K demonstrate a bigger dispersion. The latter fact reflects a specific feature of the pulse field experiments in the superfluid helium, where the experimental set-up becomes more sensitive to a mechanical noise as compared to the case of normal helium ($T$ =4.2 K). It is worth noting that the temperature measurements at ω/2π=60 GHz showed that the *g*-factor for magnetic resonance in the AFQ phase is temperature independent [22-24]. This finding agrees with the result of the present experiment, which demonstrates a good coincidence of the *g*-factor values at 1.8 K and 4.2 K for the frequency ω/2π~100 GHz (fig. 4). For higher frequencies the difference between reflectance data at 1.8 K and 4.2 K is within experimental error and therefore does not suggest any significant influence of temperature for $T<T_Q$.

Considering a frequency effect we wish to mark a tendency to the *g*-factor enhancement for the resonance A from *g*(ω/2π=44 GHz)~1.55 to *g*(ω/2π>250 GHz)~1.7 (fig. 4). Therefore in the range 44-360 GHz, where frequency vary about 8 times, the slope of the dispersion curve ω(*B* ) for the mode A is changed by 10 %. Taking into account that $CeB_6$ is a paramagnetic material, it is possible to suppose that the observed behaviour merely reflects the difference between the external and local field (the latter quantity defines the resonant frequency). However the magnetisation and susceptibility data for [110] crystallographic direction [5,23] give the magnitude of the *g*-factor change that does not exceed 3%. This estimate shows that the observed variation of the *g*-factor with frequency may be explained by the field-induced renormalization of the oscillating magnetic moments. It is natural to expect that this renormalization is a consequence of an interaction between localized magnetic moments or between localized magnetic moments and band electrons. At a moment it is not clear whether these effects can be correctly accounted in the existing models of the orbital ordering in $CeB_6$.

For the mode B, which is observed in the quasi-optical experiment only, the $g$-factor value are also tends to increase with frequency (fig. 4). Unfortunately the possible frequency and temperature effects are within experimental error, and more experimental work is required to obtain a correct $g(\omega,T)$ dependences for this magnetic resonance. Nevertheless the appearance of a new magnetic resonance mode in the high-frequency region looks very important as long as it may reflect a change of the $CeB_6$ magnetic structure in the magnetic fields exceeding ~10 T. The possible complicated structure of the orbitally ordered phase in $CeB_6$ cannot be foreseen in the antiferro-quadrupole model [8-15]. Moreover, the simultaneous observation of the orbital ordering resonance (mode A) and resonance B does not allow explaining the appearance of the latter spectral feature by a field-induced re-entrant transition into paramagnetic phase expected in the antiferro-quadrupole anzatz [10,15].

Therefore, in our opinion, an alternative model explaining low temperature magnetic ordering and orbital effects in $CeB_6$ should appear on the agenda. As a possible starting point an approach based on the ordering of the complex orbitals initially suggested by Khomskii *et al.* for manganites [25-26] may be revisited. As long as a characteristic feature of [25-26] is the transition into the orbitally ordered phase without lattice distortion, this model fails in description of manganites but exactly meets the case of $CeB_6$. Interesting that the order parameter of the model [25-26] is the magnetic octupole and consequently the excessive magnetisation responsible for the mode of the orbital ordering resonance (mode A) [23] may have octupolar nature. At present no modes of the magnetic oscillations related with the magnetic octupoles there considered theoretically, but the observation of the specific magnetic resonances in the orbitally ordered phase of $CeB_6$ shows that this direction of research may be rewarding.

In conclusion, we have successfully observed the orbital ordering resonance in $CeB_6$ in a wide frequency range 44-360 GHz using different experimental techniques. The $g$-factor for this mode is found to increase with frequency probably reflecting the renormalization of the oscillating magnetic moment. In addition to the orbital ordering resonance for the frequencies exceeding 200 GHz a new magnetic resonance with the *g*-factor 1.2-1.3 is detected.

**Acknowledgements.**

The research was carried under the frame of Cooperation Agreement between Molecular Photoscience Research Center of Kobe University, Japan and General Physics Institute of Russian Academy of Sciences. Financial support from the INTAS project 03-51-3036 the program of the Russian Academy of Sciences "Strongly Correlated Electrons" is acknowledged. This work was partly supported by Grant-in-Aid for Scientific Research on Priority Areas "High Field Spin Science in 100T " (No. 451) from the Ministry of Education, Culture, Sports, Science and Technology (MEXT) of Japan and Grant-in-Aid for Scientific Research (B) 16340106 from the Japan Society for the Promotion of Science (JSPS) and by RFBR grants 04-02-16574 and 04-02-16721.

**Figure legends.**

Fig. 1. Schema of the reflection measurements.

Fig. 2. The orbital ordering resonance (A) in cavity experiment at 1.8 K.

Fig. 3. Reflection spectra at 4.2 K for the high frequency range. The scale of the magnetic field is reduced to the scale of the *g*-factor. A- orbital ordering resonance; B- new magnetic resonance mode.

Fig. 4. The *g*-factor values for different magnetic resonances in $CeB_6$.

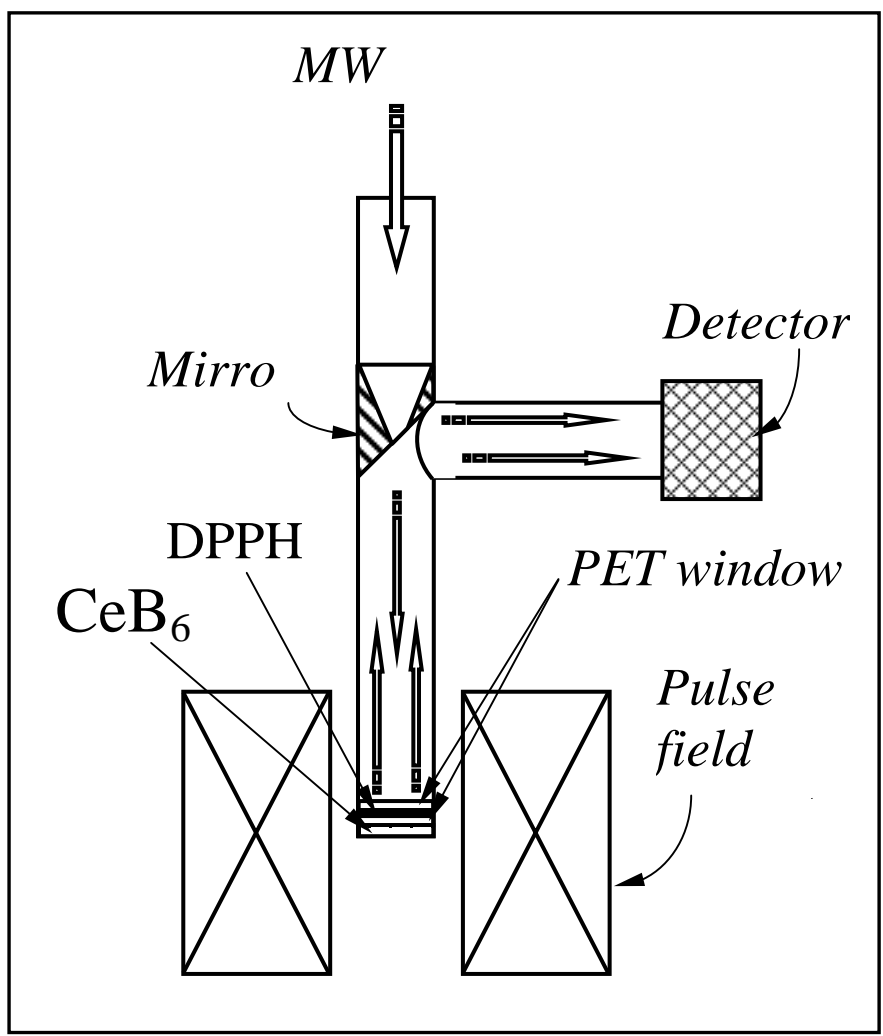

Figure 1. *S.V.Demishev et al.*

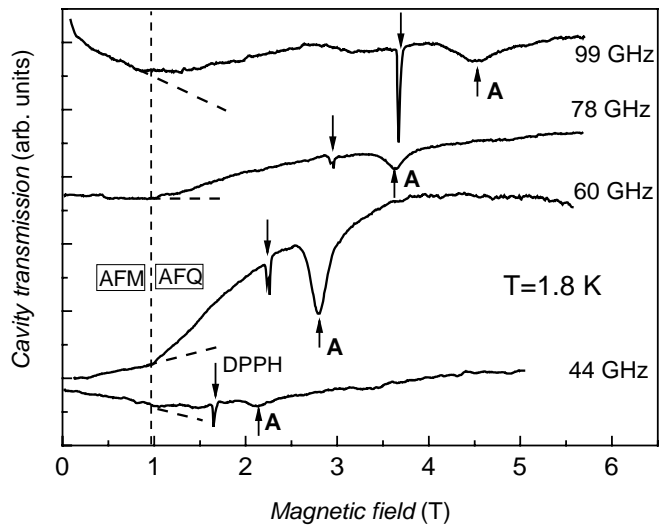

Figure 2. *S.V.Demishev et al.*

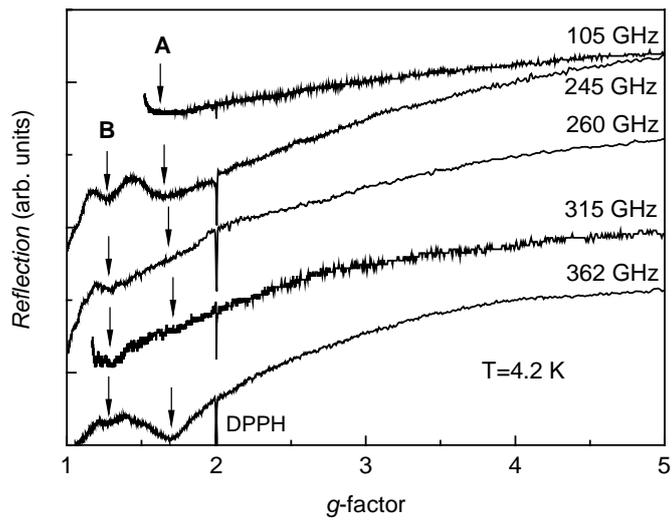

Figure 3. *S.V.Demishev et al.*

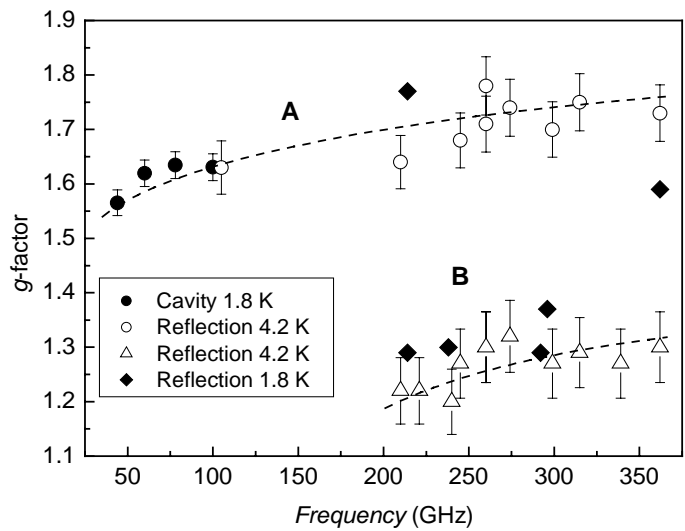

Figure 4. *S.V.Demishev et al.*